\documentclass[12pt]{article}
\usepackage[super,compress]{cite}
\usepackage{graphicx}
\usepackage{epsfig}
\usepackage{amssymb}
\begin{document}

\begin{center}
{\bf
Real part of $pp$ scattering amplitude
in Additive Quark Model
at LHC energies}

\vspace{.2cm}

Yu.M. Shabelski and A.G. Shuvaev \\

\vspace{.5cm}

Petersburg Nuclear Physics Institute, Kurchatov National
Research Center\\
Gatchina, St. Petersburg 188300, Russia\\
\vskip 0.9 truecm
E-mail: shabelsk@thd.pnpi.spb.ru\\
E-mail: shuvaev@thd.pnpi.spb.ru

\vspace{1.2cm}

\end{center}

\begin{abstract}
\noindent
Elastic $pp$ scattering at LHC energies
is treated in Additive Quark Model together with
Pomeron exchange theory.
The obtained results are compared with
the new experimental data on the ratio
of real to imaginary part of the scattering
amplitude at the small transverse momenta
\end{abstract}

\noindent
Elastic $pp$ scattering at the high energies including LHC
energies has been treated in our previous papers
\cite{Shabelski:2014yba, Shabelski:2015bba, Shabelski:2016aek}
in the framework of Additive Quark Model (AQM). The total interaction
cross section $\sigma_{tot}$, differential cross section
$d\sigma/dt$ at $\lesssim 1$~GeV$^2$ as well as its slope
$B_{pp}(t=0)$ and the ratio of the real to the imaginary part
of the amplitude,
\begin{equation}
\label{rho}
\rho\,=\,\frac{\mathrm{Re}\,A}{\mathrm{Im}\,A},
\end{equation}
have been in reasonable agreement with the experimental
data. The ratio $\rho$ (\ref{rho}) has been recently
measured by TOTEM collaboration \cite{TOTEM}.
For this reason we discuss here the value of $\rho$
that results in AQM and how it depends on the model parameters.

To begin with we briefly recall the basics of AQM.
In AQM baryon is treated as a system of
three spatially separated compact objects -- the constituent quarks.
Each constituent quark is colored and has an internal quark-gluon
structure and a finite radius that is much less than the radius of
the proton, $r_q^2 \ll r_p^2$.
The constituent quarks play the roles of incident particles
in terms of which $pp$ scattering is described in AQM.
Elastic amplitudes for large energy $s=(p_1+p_2)^2$
and small momentum transfer $t$ are dominated by Pomeron
exchange. We neglect the small difference in $pp$ and $p\bar p$
scattering coming from the exchange of negative signature
Reggeons,
Odderon (see e.g.~\cite{Avila} and references therein),
$\omega$-Reggeon etc., since their contributions
are suppressed by $s$.
The single $t$-channel exchange results into
the amplitude of constituent quarks scattering,
\begin{equation}
\label{Mqq}
A_{qq}^{(1)}(s,t) = \gamma_{qq}(t) \cdot
\left(\frac{s}{s_0}\right)^{\alpha_P(t) - 1} \cdot
\eta_P(t) \;,
\end{equation}
where $\alpha_P(t) = \alpha_P(0) + \alpha^\prime_P\cdot t$
is the Pomeron trajectory specified by the intercept
and slope values $\alpha_P(0)$ and $\alpha^\prime_P$, respectively.
The Pomeron signature factor,
$$
\eta_P(t) \,=\, i \,-\, \tan^{-1}
\left(\frac{\pi \alpha_P(t)}2\right),
$$
determines the complex structure of the amplitude. The factor
$\gamma_{qq}(t)=g_1(t)\cdot g_2(t)$ has the meaning
of the Pomeron coupling to the beam and target particles,
the functions $g_{1,2}(t)$ being the vertices of the constituent
quark-Pomeron interaction.
In the following we assume the Pomeron trajectory
to have the simplest form,
$$
\left(\frac{s}{s_0}\right)^{\alpha_P(t) - 1}\,=\,e^{\Delta\cdot\xi}
e^{-r_q^2\,q^2}, ~~ \xi\equiv \ln\frac{s}{s_0},~~
r_q^2\equiv \alpha^\prime\cdot\xi.
$$
The value $r_q^2$ defines the radius of the quark-quark interaction,
while $S_0=(9~{\rm GeV})^2$ has the meaning of typical energy scale
in Regge theory.

\begin{figure}[htb]
\centering
\vskip -1 cm
\includegraphics[width=.6\hsize]{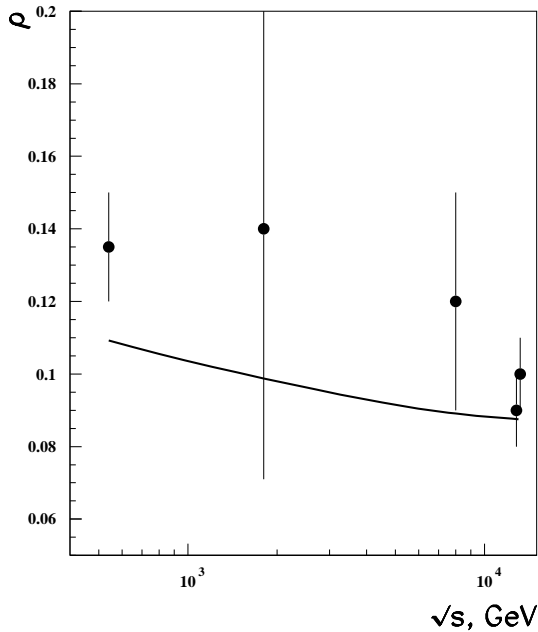}
\vskip -2 cm
\caption{\footnotesize
The energy dependence of the ratio
$\rho$.
The experimental points
are taken from \cite{Augier:1993sz, Amos:1991bp,
Antchev:2016vpy, TOTEM}.
}
\label{rhs}
\end{figure}

The scattering amplitude is presented in AQM as a sum over
the terms with a given number of Pomerons,
\begin{equation}
\label{totamp}
A_{pp}(s,t)\,=\,\sum_n A_{pp}^{(n)}(s,t),
\end{equation}
where the amplitudes $A_{pp}^{(n)}$ collect all
diagrams comprising various connections of the beam and target
quark lines with $n$ Pomerons.
Similar to Glauber theory \cite{Glaub, FG}
the multiple interactions between the same
quark pair has to be ruled out.
AQM permits the Pomeron to connect any two quark lines only once.
It crucially decreases the combinatorics, leaving the diagrams
with no more than $n=9$ effective Pomerons,
\begin{eqnarray}
\label{Mn}
A_{pp}^{(n)}(s,t)\,&=&\,i^{n-1}\biggl(\gamma_{qq}\eta_P(t_n)
e^{\Delta\cdot\xi}\biggr)^n\,
\int\frac{d^2q_1}{\pi}\cdots \frac{d^2q_n}{\pi}
\,\pi\,\delta^{(2)}(q_1+\ldots +\,q_n-Q)\,\\
&&\times\,e^{-r_q^2(q_1^2+\ldots + q_n^2)}\,
\frac 1{n!}\sum\limits_{n~\rm connections}\hspace{-1.5em}
F_P(Q_1,Q_2,Q_3)\,F_P(Q_1^{\,\prime},Q_2^{\,\prime},Q_3^{\,\prime}),
~~~t_n\simeq t/n. \nonumber
\end{eqnarray}
The sum in this formula refers to all distinct ways
to connect the beam and target quark lines with $n$ Pomerons
in the scattering diagram. The set of momenta $Q_i$ and $Q_l^{\,\prime}$
the quarks acquire from the attached Pomerons is particular
for each connection pattern.
It is worth pointing out that
the reduced combinatorics ($n\le 9$) weakens
the role of the screening correction in AQM
compared to the widely applied eikonal models.

The function $F_P(Q_i)$ plays the role of a proton form factor
for the strong interaction,
\begin{equation}
\label{FP}
F_P(Q_1,Q_2,Q_3)\,=\,\int dk_i\,\psi^*(k_1,k_2,k_3)\,
\psi(k_1+Q_1,k_2+Q_2,k_3+Q_3).
\end{equation}
Here $\psi(k_1,k_2,k_3)$,
is the initial proton wave function in terms of the quarks'
transverse momenta $k_i$, while
$\psi(k_i+Q_i)\equiv \psi(k_1+Q_1,k_2+Q_2,k_3+Q_3)$
is the wavefunction of the scattered proton.
A more detailed description can be found
in~Ref.~\cite{Shabelski:2014yba}.

The quarks' wave function has been taken in the simple form
of Gaussian packets,
\begin{equation}
\label{gausspack}
\psi(k_1,k_2,k_3)\,=\,N\bigl[\,e^{-a_1(k_1^2+k_2^2+k_3^2)}\,
+\,C_1\,e^{-a_2(k_1^2+k_2^2+k_3^2)}
+\,C_2\,e^{-a_3(k_1^2+k_2^2+k_3^2)}\bigr],
\end{equation}
normalized to unity.
The parameters have been chosen
in \cite{Shabelski:2016aek} to fit the
of $d\sigma/dt$ distribution, namely the
slope at $t=0$ and the position of the minimum
evidently seen in the experimental data
for $\sqrt{s}=7$~TeV \cite{TO1,TO2}.
They read
\begin{equation}
\label{param}
\Delta=0.14,~~~
\alpha^\prime=0.116\,{\rm GeV}^{-2},~~~
\gamma_{qq}=0.45\,{\rm GeV}^{-2}.
\end{equation}
$$
a_1=9.0\,{\rm GeV}^{-2},~~~
a_2=0.29\,{\rm GeV}^{-2},~~~
a_3=2.0\,{\rm GeV}^{-2},~~~
C_1=0.024,~~~
C_2=0.05.
$$

Here the same set of parameters is utilized to compare the energy
behavior of $\rho$ (\ref{rho}) to the experimental data
available now up to $\sqrt s = 13$~TeV, Fig.\ref{rhs}.
The new measurement presented by TOTEM \cite{Antchev}
gives two different $\rho$ values
at $\sqrt s = 13$~TeV
(connected to different assumptions
made for the data analysis)
shown in Fig.\ref{rhs}.
Though our curve generally
passes a little below the data it correctly
reproduces the overall energy trend.

\begin{figure}[htb]
\includegraphics[width=.45\textwidth]{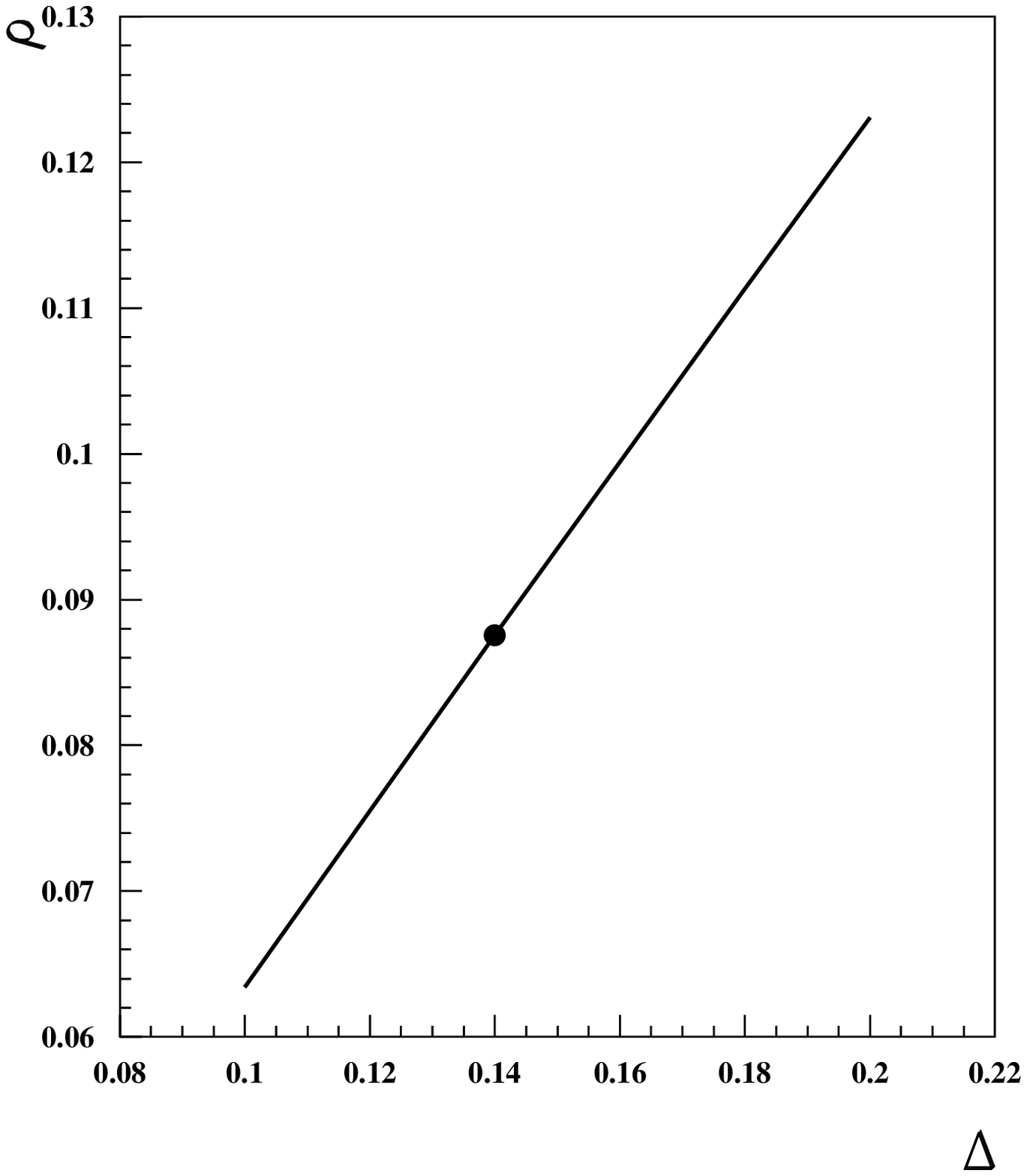}
\includegraphics[width=.45\textwidth]{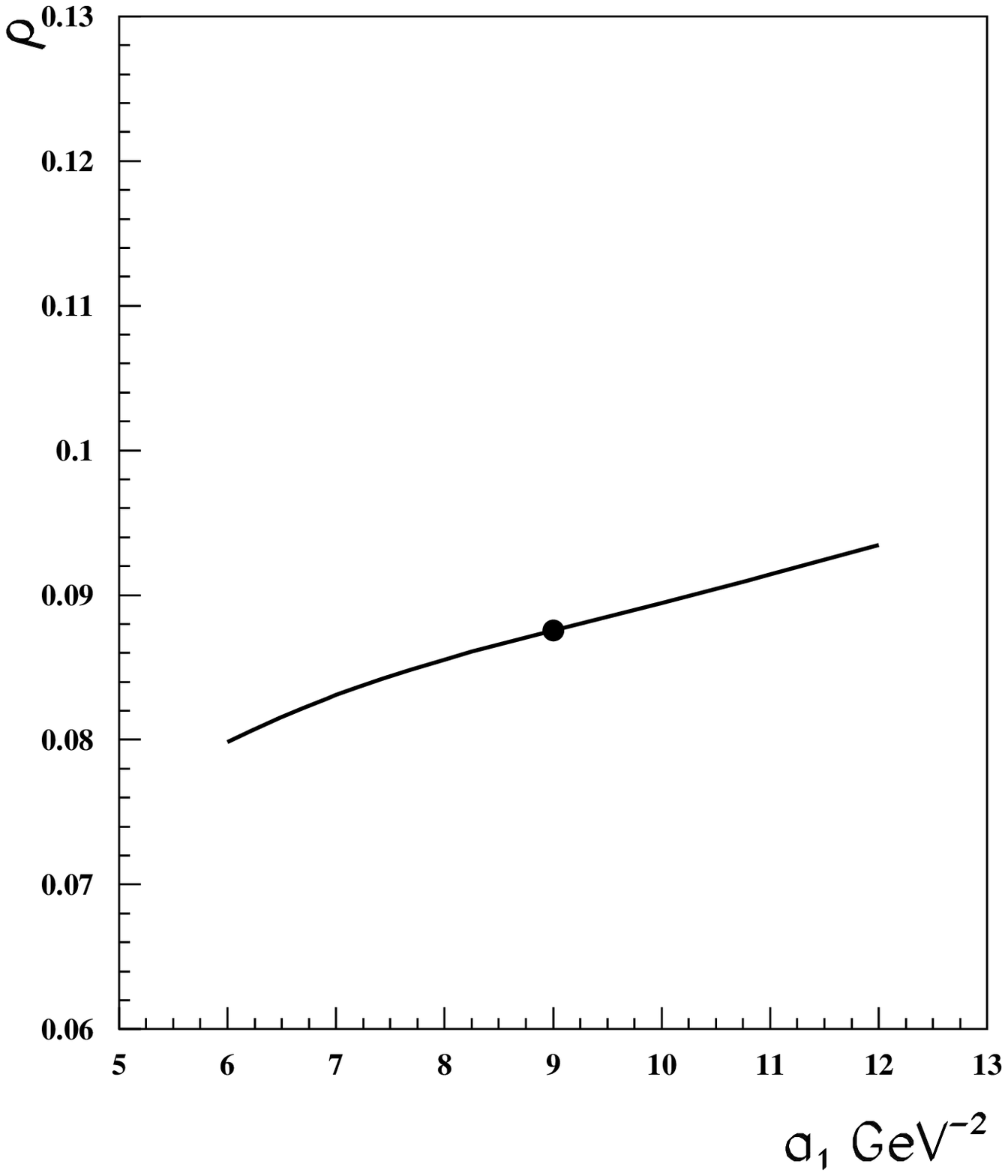}
\vskip -2 cm
\caption{\footnotesize
The ratio $\rho$ as a function of parameter $\Delta$
(left) and as a function of parameter $a_1$
(right).
The filled points indicate the parameters fixed
in the set (\ref{param}).}
\label{rhda}
\end{figure}

To demonstrate how the input parameters affect
the result, the ratio $\rho$ is presented
in Fig.\ref{rhda} (left) versus the intercept $\Delta$,
i.e. the shift of bare Pomeron from the unity.
The parameter $\gamma_{qq}$ is varied along with
$\Delta$ to keep the fixed cross section
$d\sigma/dt$ at $t=0$. The value $\rho$ is seen
to substantially increase when $\Delta$ growths
while another observables like $\sigma_{tot}$,
$B_{pp}$ etc remain unchanged.
Similarly, Fig.\ref{rhda} (right)
shows $\rho$
as a function of parameter $a_1$, which
determines the largest range of the quark-quark
interaction in (\ref{gausspack}).
Another physical quantities are again fixed
to the values they have near $t=0$.
The ratio $\rho$ increases with the growth of $a_1$.

To summarize, our approach yields the value $\rho$
(\ref{rho}) a little below the data but with
the correct energy behavior. It is possibly related
to the rather small absorptive corrections for
the elastic scattering $pp$ amplitude. Similarly to
Ref.\cite{Khoze:2017swe} any odderon contribution
are not needed. It was shown that small change
of the evaluated $\rho$ is achieved by the variation
of model parameters without affecting another
observable quantities at $t \approx 0$.


\begin{thebibliography}{**}
\bibitem{Shabelski:2014yba}
Y.~M.~Shabelski and A.~G.~Shuvaev,
JHEP {\bf 1411} (2014) 023
[arXiv:1406.1421 [hep-ph]].

\bibitem{Shabelski:2015bba}
Y.~M.~Shabelski and A.~G.~Shuvaev,
Eur.\ Phys.\ J.\ C {\bf 75} (2015) 9,  438
[arXiv:1504.03499 [hep-ph]].

\bibitem{Shabelski:2016aek}
Y.~M.~Shabelski and A.~G.~Shuvaev,
Eur.\ Phys.\ J.\ C {\bf 76} (2016) no.8,  470
[arXiv:1601.04426 [hep-ph]].

\bibitem{TOTEM}
TOTEM Collaboration, 132nd LHCC open session, CERN, Nov. 30th 2017.

\bibitem{Avila}
R.~Avila, P.~Gauron and B.~Nicolescu,
Eur.\ Phys.\ J.\ C {\bf 49}, 581 (2007) [hep-ph/0607089].

\bibitem{Glaub} R.~J.~Glauber. In "Lectures in Theoretical Physics",
Eds. W.~E.~Brittin etal., New York (1959), vol.1, p.315.

\bibitem{FG} V.~Franco and R.~J.~Glauber, Phys.Rev. {\bf 142}
(1966) 1195.

\bibitem{TO1} TOTEM Collaboration, G. Antchev et al.,
Europhys.Lett. {\bf 101} (2013) 21002.

\bibitem{TO2} TOTEM Collaboration, G. Antchev et al.,
Europhys.Lett. {\bf 95} (2011) 41001, [arXiv:1110.1385].

\bibitem{Antchev}
G. Antchev et al. (TOTEM collaboration), no. CERN-EP-2017-335, (2017).\\
https://cds.cern.ch/record/2298154.

\bibitem{Augier:1993sz}
C.~Augier {\it et al.} [UA4/2 Collaboration],
Phys.\ Lett.\ B {\bf 316} (1993) 448.

\bibitem{Amos:1991bp}
N.~A.~Amos {\it et al.} [E710 Collaboration],
Phys.\ Rev.\ Lett.\  {\bf 68} (1992) 2433.

\bibitem{Antchev:2016vpy}
G.~Antchev {\it et al.} [TOTEM Collaboration],
Eur.\ Phys.\ J.\ C {\bf 76} (2016) no.12,  661
[arXiv:1610.00603 [nucl-ex]].


\bibitem{Khoze:2017swe}
V.~A.~Khoze, A.~D.~Martin and M.~G.~Ryskin,
arXiv:1712.00325 [hep-ph].
\end{thebibliography}
\end{document}